\documentclass[prl,aps,superscriptaddress,showpacs,twocolumn]{revtex4}

\usepackage{bbm}
\usepackage{amsmath}
\usepackage{amssymb}
\usepackage[dvips]{graphicx}
\usepackage{type1cm}
\DeclareGraphicsExtensions{.ps}

\newcommand{\dicke}[2]{|D_{#1}^{(#2)}\rangle}

\begin{document}

\title{Operational Families of Entanglement Classes for Symmetric $N$-Qubit States}

\date{\today}

\author{T. Bastin}
\affiliation{Institut de Physique Nucl\'eaire, Atomique et de
Spectroscopie, Universit\'e de Li\`ege, 4000 Li\`ege, Belgium}

\author{S. Krins}
\affiliation{Institut de Physique Nucl\'eaire, Atomique et de
Spectroscopie, Universit\'e de Li\`ege, 4000 Li\`ege, Belgium}

\author{P. Mathonet}
\affiliation{Institut de Math\'ematique, Universit\'e de Li\`ege, 4000 Li\`ege, Belgium}

\author{M. Godefroid}
\affiliation{Chimie quantique et Photophysique, CP160/09, Universit\'e Libre de Bruxelles, 1050 Bruxelles,
Belgium}

\author{L. Lamata}
\affiliation{Max-Planck-Institut f\"ur Quantenoptik,
Hans-Kopfermann-Strasse 1, 85748 Garching, Germany}

\author{E. Solano}
\affiliation{Departamento de Qu\'{\i}mica F\'{\i}sica, Universidad del Pa\'{\i}s Vasco - Euskal Herriko Unibertsitatea, Apdo.\ 644, 48080 Bilbao, Spain}
\affiliation{IKERBASQUE, Basque Foundation for Science, Alameda Urquijo 36, 48011 Bilbao, Spain}

\begin{abstract}
We solve the entanglement classification under stochastic local operations and classical communication (SLOCC) for all multipartite symmetric states in the general $N$-qubit case. For this purpose, we introduce 2 parameters playing a crucial role, namely the \emph{diversity degree} and the \emph{degeneracy configuration} of a symmetric state. Those parameters give rise to a simple method of identifying operational families of SLOCC entanglement classes of all symmetric $N$-qubit states, where the number of families grows as the partition function of the number of qubits.
\end{abstract}


\maketitle

Entanglement is considered as a key resource
in quantum information. Many tasks which are unpractical or even impossible with classical schemes can be made possible with entangled states~\cite{Nielsen}. For multipartite systems entanglement can be of different kinds~\cite{Dur00} and its general classification represents a difficult challenge. Entanglement classes that are inequivalent under stochastic local operations and classical
communication (SLOCC) are of fundamental importance because states belonging to each class can be used for similar quantum information tasks. Consequently, the development of a simple characterization for identifying SLOCC classes is highly desirable.
To date this question remains unsolved in the general multipartite case, as the algebraic complexity increases very rapidly with the number of parties. For $N$-qubit systems, the number of entanglement
classes is finite only for $N \leqslant 3$~\cite{Dur00}. For $N=2$, only 2 SLOCC classes exist, including the separable (S) class. For $N = 3$, 6 entanglement classes exist~\cite{Dur00}, among
them 3 contain symmetric states~: the Greenberger-Horne-Zeilinger (GHZ), W, and S classes. For $N=4$ the number of SLOCC {\it classes} is infinite~\cite{Dur00},
and it makes sense to try to gather those having some similar properties into a finite number of \emph{families}. In this sense, different methods, based on specific mathematical properties, have been proposed so far~\cite{Ver03,Ost05,Lam07}.
For $N \geqslant 5$, only particular features are explicitly known. Though the inductive method of Lamata {\it et
al.}~\cite{Lam07} may provide a solution, the number of entanglement families is
expected to grow exponentially with the number of qubits, being at the level of hundreds for $N=5$. It would be certainly desirable that the families of entangled SLOCC classes are finite, grow slowly with the number of qubits, and, most importantly, that they are physically meaningful. Recently, a key step has been done relating entanglement classes with specific physical setups, unveiling an operational approach to entanglement classification~\cite{Bas09}. Furthermore, impressive experiments and analysis have been performed in linear optics setups~\cite{Wieczorek0809}.

In this Letter, we solve the problem of SLOCC class identification for all multipartite symmetric states~\cite{Sto03} in the general $N$-qubit case. To this end, we introduce 2 key parameters, namely the \emph{diversity degree} and the \emph{degeneracy configuration} of a symmetric state. Those parameters give rise to an easy identification method and allow us to identify the SLOCC classes with a complexity growing slowly with low numbers of qubits.
For example, for
$N=4, 5$ and $6$, we find $5$, $7$, and $11$ entanglement families,
respectively. This result is not only another attractive manner of building families of SLOCC entanglement classes based on a distinctive mathematical property. It also provides operational families classifying and experimentally generating all multipartite
symmetric entangled states of $N$ qubits with a straightforward experimental implementation. In this respect,
applications to 
nonlocal tests with many qubits~\cite{Mer90}, quantum metrology~\cite{Gio04}
or quantum algorithms~\cite{Nielsen} could be performed and studied under an operational scope.

Two $N$-partite states are said to belong to the same SLOCC entanglement
class if they can be obtained from each other via
local operations and classical communication with a certain
probability. Mathematically speaking, two states belong to the same
class if and only if they are connected via an invertible local
operation (ILO)~\cite{Dur00}~:
\begin{equation}
    |\psi\rangle, |\phi\rangle \in \textrm{ same class } \:\: \Leftrightarrow \:\: |\psi\rangle = A_1 \otimes \cdots \otimes A_N |\phi\rangle,
\end{equation}
where $A_i$ ($i = 1, \ldots, N$) represent ILOs ($\det A_i
\neq 0$). When two states $|\psi_S\rangle$ and $|\phi_S\rangle$ are symmetric
with respect to the permutations of the parties, it is sufficient to look
for a symmetric ILO~\cite{Mat09}:
\begin{equation}
    |\psi_S\rangle, |\phi_S\rangle \in \textrm{ same class } \:\: \Leftrightarrow \:\: |\psi_S\rangle = A \otimes \cdots \otimes A |\phi_S\rangle,
\end{equation}
with the same ILO acting on each qubit.

{\it{Diversity degree and degeneracy configuration of a symmetric state.--}}
In an $N$-qubit system, any symmetric state $|\psi_S\rangle$ can be written in the form~\cite{Maj32}
\begin{equation}
\label{psiS} |\psi_S\rangle = \mathcal{N} \!\!\!\! \sum_{1 \leqslant i_1 \neq
\ldots \neq i_N \leqslant N} |\epsilon_{i_1}, \ldots,
\epsilon_{i_N}\rangle,
\end{equation}
where $\mathcal{N}$ is a normalization prefactor and the $|\epsilon_i\rangle$'s are single qubit states $\alpha_i |1\rangle + \beta_i
|0\rangle$ with $|\alpha_i|^2 + |\beta_i|^2 = 1$. By using Vieta's formulas~\cite{Bas09}, Eq.~(\ref{psiS}) can be obtained from the expansion
\begin{equation}
\label{psiS2} |\psi_S\rangle = \sum_{k=0}^N d_k \dicke{N}{k},
\end{equation}
with $\dicke{N}{k}$ the symmetric Dicke states with $k$
$|0\rangle$ components and $d_k$ complex coefficients. Eq.~(\ref{psiS2}) reduces to Eq.~(\ref{psiS}) up to an insignificant global phase factor when the $|\epsilon_i\rangle$ states are defined with $\alpha_i/\beta_i$ equal to
 the $K$ roots of the polynomial $P(z) = \sum_k^N (-1)^{k}
\sqrt{C_N^k} d_k z^k$, being $K$ the polynomial degree and
the remaining $\alpha_i$ equal to 1. Here $C_N^k$ denotes the
binomial coefficient of $N$ and $k$.

In Eq.~(\ref{psiS}), several $|\epsilon_i\rangle$ states may be identical and we call their number the \emph{degeneracy number}. Up to an
insignificant global phase factor, two states $|\epsilon_i\rangle$
and $|\epsilon_j\rangle$ are identical if and only if $\alpha_i
\beta_j - \alpha_j \beta_i = 0$.
We then define the \emph{degeneracy configuration}
$\mathcal{D}_{\{n_i\}}$ of a symmetric state $|\psi_S\rangle$ as the list of
its degeneracy numbers $n_i$ ordered by convention in decreasing order. The dimension of this list (number of $n_i$'s) defines the \emph{diversity degree} $d$ of the symmetric state. It represents merely the number of \emph{distinct} $|\epsilon_i\rangle$ states in Eq.~(\ref{psiS}).
As we have $\sum_i n_i = N$, the degeneracy configuration is nothing
else than a partition of $N$ and the number of different degeneracy
configurations for an $N$-qubit system is simply given by the partition function $p(N)$~\cite{footnote1}.
For example, a symmetric $N$-qubit state with all $\epsilon_i$
identical has a degeneracy configuration $\mathcal{D}_N$ and a
diversity degree $d$ of 1. If all but one $\epsilon_i$ are
identical, we get the configuration $\mathcal{D}_{N-1,1}$ and $d=2$.
If all but two $\epsilon_i$ are identical, we get the configuration
$\mathcal{D}_{N-2,2}$ ($d=2$) or $\mathcal{D}_{N-2,1,1}$ ($d=3$)
depending on whether the two remaining ones are identical or not,
respectively. If all $\epsilon_i$ are distinct we get the
configuration $\mathcal{D}_{1,\ldots,1}$ ($d=N$). In a 2-qubit
system, there are only 2 different degeneracy configurations, namely
$\mathcal{D}_{2}$ and $\mathcal{D}_{1,1}$~: $p(2) = 2$. For the lowest $N$'s, we
have notably $p(3)=3$, $p(4)=5$, $p(5)=7$, $p(6)=11$,
$p(7)=15$, $p(8) = 22$, $p(9) = 30$, and $p(10) = 42$. For large $N$, $p(N)$ scales as $\exp({\pi \sqrt{2 N / 3}})/4 N \sqrt{3}$.
In the general case $N$, there is a unique degeneracy configuration
having $d=1$, namely $\mathcal{D}_N$, and $d=N$, namely $\mathcal{D}_{1,\ldots, 1}$. For $d=2$, we get $N/2$
[$(N-1)/2$] degeneracy configurations if $N$ is even (resp.\
odd), namely $\mathcal{D}_{N-1,1}, \ldots, \mathcal{D}_{N/2,N/2}$ [$\mathcal{D}_{N-1,1}, \ldots,
\mathcal{D}_{(N+1)/2,(N-1)/2}$].

{\it{SLOCC action on symmetric states.--}} The action of a symmetric
ILO $A^{\otimes N}$ on an arbitrary symmetric state $|\psi_S\rangle$ written in the form of Eq.~(\ref{psiS}) reads
\begin{equation}
|\psi'_S\rangle = \mathcal{N} \sum_{1 \leqslant i_1 \neq \ldots \neq i_N \leqslant
N} |\epsilon'_{i_1}, \ldots, \epsilon'_{i_N}\rangle,
\end{equation}
with
\begin{equation}
    \label{epspi}
    |\epsilon'_i\rangle = A |\epsilon_i\rangle, \quad i = 1, \ldots, N.
\end{equation}

In other words, two symmetric states
$|\psi_S\rangle = \mathcal{N} \sum |\epsilon_{i_1}, \ldots, \epsilon_{i_N}\rangle$
and
$|\psi'_S\rangle = \mathcal{N}' \sum |\epsilon'_{i_1}, \ldots, \epsilon'_{i_N}\rangle$
belong to the same SLOCC class if and only if there exists a
\emph{single} invertible local operation $A$ able to convert
\emph{each} $|\epsilon_i\rangle$ state into \emph{each} of the
$|\epsilon'_i\rangle$ states, possibly up to a multiplicative constant. Hence the degeneracy configuration of
an arbitrary state $|\psi_S\rangle$ is \emph{invariant} under SLOCC
operations since those operations convert \emph{all} $\epsilon_i$ in
a \emph{similar} way. Thus when two symmetric states
differ in their degeneracy configuration, they belong necessarily to
different SLOCC classes. The converse is not necessarily true~: two states with
the same degeneracy configuration do not belong necessarily to the
same SLOCC class. For this to hold, we would have to find a local
operation $A$ able to convert all $\epsilon_i$ of the first state
into all $\epsilon'_i$ of the second state, which is not guaranteed when the two states differ in more than 3 $\epsilon_i$'s as shown hereafter explicitly.

{\it{Families of symmetric SLOCC entanglement classes.--}} The
invariance of the degeneracy configuration with respect to
SLOCC for symmetric states suggests that for an $N$-qubit system we can define
$p(N)$ families of SLOCC entanglement classes, each of them gathering all symmetric states with the same degeneracy configuration. In this way, we ensure
that states belonging to different families are indeed SLOCC
incompatible. We denote those families by their degeneracy
configuration and by the diversity degree of the states they gather. We can state then the important result~: all families with a diversity degree $d$ lower or
equal than 3 gather states belonging to a single SLOCC class. All
other families contains an infinite number of SLOCC classes (see demonstration below).

{\it{$d=1$: $\mathcal{S}$ family}.--} For $d = 1$, the only family is
$\mathcal{D}_N \equiv \mathcal{S}$. This family gathers all states of the form
\begin{equation}
|\psi_{\mathrm{sep}}\rangle = |\epsilon, \ldots, \epsilon\rangle,
\end{equation}
with $|\epsilon\rangle = \alpha |1\rangle + \beta |0\rangle$.
Considering an invertible local operation $A$ transforming the
$|\epsilon\rangle$ state into the $|\lambda\rangle \equiv |1\rangle$
state, namely $|1\rangle = A |\epsilon\rangle,$ all states of the
$\mathcal{D}_N$ family can always be converted via an ILO to the
canonical state $|S_N\rangle = |1, \ldots, 1\rangle$. Hence, the
$\mathcal{D}_N$ family identifies to the only separable state class for symmetric
states.

{\it{$d = 2$: $\mathcal{W}$ families.--}} The case $d=2$ contains all families
$\mathcal{D}_{N-k,k}$ for $k = 1, \ldots, N/2$ if $N$ is even ($k =
1, \ldots, (N-1)/2$ if $N$ is odd). Any of those families gathers
states of the form
\begin{equation}
    |\psi_W\rangle = \mathcal{N} \sum_{P(\epsilon_1, \epsilon_2)} |\underbrace{\epsilon_1, \ldots, \epsilon_1}_{N-k}, \underbrace{\epsilon_2, \ldots, \epsilon_2}_{k}\rangle,
\end{equation}
with $|\epsilon_1\rangle \neq |\epsilon_2\rangle$ and where
$P(\epsilon_1, \epsilon_2)$ stands for all $C_N^k$ permutations of
$\epsilon_1$ and $\epsilon_2$ appearing $N-k$ and $k$ times, respectively. Considering an invertible local
operation $A$ transforming $|\epsilon_1\rangle$ into
$|\lambda_1\rangle \equiv |1\rangle$ and $|\epsilon_2\rangle$ into
$|\lambda_2\rangle \equiv |0\rangle$, namely
\begin{equation}
\label{ILO2}
    |1\rangle = A |\epsilon_1\rangle, \quad |0\rangle = A |\epsilon_2\rangle,
\end{equation}
all states of a $\mathcal{D}_{N-k,k}$ family can always be converted
into the canonical state
\begin{equation}
    \dicke{N}{k} \equiv \frac{1}{\sqrt{C_N^k}} \sum_{P(1,0)} |\underbrace{1, \ldots, 1}_{N-k}, \underbrace{0, \ldots, 0}_{k}\rangle.
\end{equation}
Hence, we find in each of the $N/2$, or $(N-1)/2$, $d = 2$ families
$\mathcal{D}_{N-k,k}$ a unique SLOCC class where the symmetric Dicke
states $\dicke{N}{k}$ appear naturally as their canonical
states, denoting these families by
\begin{equation}
    \mathcal{W}_k \equiv \mathcal{D}_{N-k,k}, \quad k = 1, \ldots N/2 \textrm{ [or  $(N-1)/2$]}.
\end{equation}
We so deduce that all symmetric Dicke states
$\dicke{N}{k}$ [$k = 1,\ldots,N/2$ [or $(N-1)/2$]] are SLOCC
incompatible~\cite{Li08}, while $\dicke{N}{k}$ and $\dicke{N}{N-k}$ belongs
to the same SLOCC class as they both belong to the family
$\mathcal{W}_k$.

{\it{$d = 3$ families.--}} The case $d=3$ contains all families
$\mathcal{D}_{n_1,n_2,n_3}$ with $n_1 + n_2 + n_3 = N$. Any of those
families gathers states of the form
\begin{equation}
\label{psid3}
    |\psi\rangle = \mathcal{N} \sum_{P(\epsilon_1, \epsilon_2, \epsilon_3)} |\underbrace{\epsilon_1, \ldots, \epsilon_1}_{n_1}, \underbrace{\epsilon_2, \ldots, \epsilon_2}_{n_2}, \underbrace{\epsilon_3, \ldots, \epsilon_3}_{n_3}\rangle,
\end{equation}
with $|\epsilon_1\rangle \neq |\epsilon_2\rangle \neq
|\epsilon_3\rangle$ and where $P(\epsilon_1, \epsilon_2,
\epsilon_3)$ stands for all permutations of $\epsilon_1$,
$\epsilon_2$ and $\epsilon_3$ appearing $n_1$, $n_2$ and $n_3$ times, respectively.
In contrast to the $d=1$ and $d=2$ case, it is not possible to consider an invertible local
operation $A$ that would transform all three $|\epsilon_i\rangle$ states
into given $|\lambda_i\rangle$ states, namely
\begin{equation}
\label{ILO3}
    |\lambda_1\rangle = A |\epsilon_1\rangle, \quad |\lambda_ 2\rangle = A |\epsilon_2\rangle, \quad |\lambda_3\rangle = A
    |\epsilon_3\rangle.
\end{equation}
Indeed, the local operation $A$ is
acting in a space of dimension 2 and Eq.~(\ref{ILO3}) is a set of 3
independent constraints impossible to fulfill simultaneously.
However, as long as we consider all $|\lambda_i\rangle$ distinct (as
are the $|\epsilon_i\rangle$ states), it is fortunately possible to
find an invertible local operation $A$ and complex numbers $e$, $f$
and $g$ so as
\begin{align}
\label{ILO3b}
    e |\lambda_1\rangle = A |\epsilon_1\rangle, \quad f |\lambda_2\rangle & = A |\epsilon_2\rangle, \quad g |\lambda_3\rangle = A |\epsilon_3\rangle, \nonumber \\
    e^{n_1} f^{n_2} g^{n_3} & = 1.
\end{align}
This equation represents a set of 7 equations for 7 unknowns (the 4
matrix elements of $A$ and the 3 numbers $e$, $f$ and $g$). The last equation makes the
system nonlinear and, however, it admits always a
solution. In this way,
all states (\ref{psid3}) of a $d=3$
family can be transformed via an ILO $A^{\otimes N}$ into any desired state
\begin{equation}
    \label{csd3}
    |\psi_c\rangle = \mathcal{N}' \sum_{P(\lambda_1, \lambda_2, \lambda_3)} |\underbrace{\lambda_1, \ldots, \lambda_1}_{n_1}, \underbrace{\lambda_2, \ldots, \lambda_2}_{n_2}, \underbrace{\lambda_3, \ldots, \lambda_3}_{n_3}\rangle,
\end{equation}
that can act as a canonical representative state of the family.
The choice of this state within the family is strictly arbitrary.
All this proves that each $d=3$ family gathers states belonging to a
single SLOCC class.

{\it{$d = 4$ families.--}}
The case $d=4$ contains all families $\mathcal{D}_{n_1,n_2,n_3,n_4}$
with $n_1 + n_2 + n_3 + n_4 = N$. Any of those families gathers
states of the form
\begin{equation}
\label{psid4}
    |\psi\rangle = \mathcal{N} \sum_{P(\epsilon_1, \ldots, \epsilon_4)} |\underbrace{\epsilon_1, \ldots, \epsilon_1}_{n_1}, \ldots, \underbrace{\epsilon_4, \ldots, \epsilon_4}_{n_4}\rangle,
\end{equation}
where $|\epsilon_1\rangle \neq \ldots \neq |\epsilon_4\rangle$ and
where $P(\epsilon_1, \ldots, \epsilon_4)$ stands for all
permutations of $\epsilon_1, \ldots, \epsilon_4$ appearing $n_1, \ldots, n_4$ times, respectively.
In this case it is unfortunately impossible to get an ILO that would
transform the state (\ref{psid4}) into any desired canonical state of the same
structure, namely
\begin{equation}
    |\psi_c\rangle = \mathcal{N}' \sum_{P(\lambda_1, \ldots, \lambda_4)} |\underbrace{\lambda_1, \ldots, \lambda_1}_{n_1}, \ldots, \underbrace{\lambda_4, \ldots, \lambda_4}_{n_4}\rangle.
\end{equation}
Indeed, for this to hold we would have to find an ILO and 4 numbers
$e$, $f$, $g$ and $h$ satisfying the conditions
\begin{align}
\label{ILO4}
    e |\lambda_1\rangle & = A |\epsilon_1\rangle, \quad f |\lambda_2\rangle = A |\epsilon_2\rangle, \nonumber \\
    g |\lambda_3\rangle & = A |\epsilon_3\rangle, \quad h |\lambda_4\rangle = A |\epsilon_4\rangle, \\
    & e^{n_1} f^{n_2} g^{n_3} h^{n_4} = 1. \nonumber
\end{align}
This system is a set of 9 equations for 8 unknowns and is thus
overdetermined. It does not admit any solution in the general case,
unless we leave a free parameter $\mu$ in the
$|\lambda_i\rangle$ states to add an unknown and remove the overdeterminacy. This shows that the states in any of the
$d=4$ families cannot be converted via an ILO to any desired
canonical state of the family. Those families
contain rather a continuous range of SLOCC classes with canonical form states $|\psi_c(\mu)\rangle$ depending on a continuous $\mu$ parameter. States having inequivalent $\mu$ parameters~\cite{footnote2} are not SLOCC convertible into each other.

{\it{$4 < d < N$ families.--}} All reasoning of the previous section
can be repeated for those families. They contain states belonging to
a continuous range of SLOCC classes. The number $n_{\mu}$ of continuous
parameters identifying the canonical representations of those states
is simply given by the difference between the equation number $2d + 1$ of the non-linear system generalizing (\ref{ILO4}) and the equation unknown number $d + 4$, that is $n_{\mu} = d - 3$. It increases only linearly with the diversity degree $d$.

{\it{$d = N$ family.--}} For $d = N$, the only family is
$\mathcal{D}_{1,\ldots, 1}$. This family gathers states of the
general form
\begin{equation}
\label{psidN} |\psi_S\rangle = \mathcal{N} \sum_{1 \leqslant i_1
\neq \ldots \neq i_N \leqslant N} |\epsilon_{i_1}, \ldots,
\epsilon_{i_N}\rangle,
\end{equation}
where all $|\epsilon_i\rangle$ individual states are distinct. For
$N > 3$, this family contains necessarily a continuous range of SLOCC
classes parametrized with $N-3$ continuous parameters. We have also
the certitude that the $\mathcal{D}_{1,\ldots, 1}$ family contains
the class of states SLOCC equivalent to
\begin{equation}
|\textrm{GHZ}_N\rangle = \frac{1}{\sqrt{2}}(|1,\ldots,1\rangle +
    |0,\ldots,0\rangle).
\end{equation}
This arises from the fact that the GHZ state can also be written in
the form (\ref{psidN}) with all distinct $|\epsilon_i\rangle$ given
by
\begin{equation}
    |\epsilon_i\rangle = \frac{1}{\sqrt{2}}(e^{-i \theta_i} |1\rangle + e^{i \theta_i} |0\rangle), \quad i = 1, \ldots, N,
 \end{equation}
where
\begin{equation}
    \theta_i = \left[ \frac{\pi}{2 N} \right]_{N \textrm{ even}} + (i - 1) \frac{\pi}{N},
\end{equation}
with the term inside the square brackets only present for the
case of an even number $N$.

We now exemplify these results with
$N=2,3$ and $4$. For higher $N$'s, the identification process extrapolates straightforwardly. For $N=2$, we have the 2 families $\mathcal{D}_2$ ($d =
1$) and $\mathcal{D}_{1,1}$ ($d = 2$). The two families have a
diversity degree $d$ lower than $4$ and thus gather states belonging
to a single SLOCC class. The $\mathcal{D}_2$ family identifies to
the separable state class for symmetric states. The
$\mathcal{D}_{1,1}$ family identifies to the entangled state class for symmetric states.

For $N = 3$, we have the 3 families
$\mathcal{D}_3$ ($d = 1$), $\mathcal{D}_{2,1}$ ($d = 2$) and
$\mathcal{D}_{1,1,1}$ ($d = 3$). These families have a
diversity degree $d < 4$ and thus gather states belonging
to a single SLOCC class. The $\mathcal{D}_3$ family identifies to
the symmetric $S$ class, the
$\mathcal{D}_{2,1}$ family to the symmetric $W$ class, and the $\mathcal{D}_{1,1,1}$ family identifies to the symmetric GHZ class.

For $N = 4$, we have the 5 families
$\mathcal{D}_4$ ($d = 1$), $\mathcal{D}_{3,1}$ ($d = 2$),
$\mathcal{D}_{2,2}$ ($d = 2$), $\mathcal{D}_{2,1,1}$ ($d=3$) and
$\mathcal{D}_{1,1,1,1}$ ($d = 4$). With the exception of the last
family, all others have a diversity degree $d$ lower than
$4$ and thus gather states belonging to a single SLOCC class. The
$\mathcal{D}_4$ family identifies to the separable state class for
symmetric states with $\dicke{4}{0}$ as a canonical
representative state. The $\mathcal{D}_{3,1}$ family identifies to the
$W_1$ class for symmetric states with $\dicke{4}{1}$ as a
canonical representative state.
The $\mathcal{D}_{2,2}$ family identifies to the $W_2$ class for
symmetric states with $\dicke{4}{2}$ as a canonical
representative state.
The $\mathcal{D}_{2,1,1}$ family identifies to the single SLOCC class of
states equivalent to the $(\dicke{4}{0} + \dicke{4}{2})/\sqrt{2}$
state.
The $\mathcal{D}_{1,1,1,1}$ family gathers states belonging to a continuous range of SLOCC classes.
All states of this family can be transformed via an ILO to the canonical form
$(|\textrm{GHZ}_4\rangle + \mu \dicke{4}{2})/\sqrt{1 + |\mu|^2}$ with complex $\mu \neq \pm 1/\sqrt{3}$.
States of different $\mu$ and varying by more than just the sign of $\mu$ are inequivalent with respect to SLOCC.


{\it{Operational families in the lab.--}}
The $p(N)$ entanglement families of symmetric states, defined on
the basis of degeneracy configurations, can be associated in a one-to-one correspondence with particular experimental parameter configurations.
In the experimental setup proposed in Ref.~\cite{Bas09}, multiqubit symmetric atomic states are produced by projective measurements of photon polarization states using photon detectors equipped with elliptical polarization filters. There, a final atomic state of the form of Eq.~(\ref{psiS}) is produced, and each single qubit state $|\epsilon_i\rangle = \alpha_i |1\rangle + \beta_i |0\rangle$ is directly determined by the polarization state $\boldsymbol{\epsilon}_i = \alpha_i \boldsymbol{\sigma}_+ + \beta_i \boldsymbol{\sigma}_-$ of the $i$th polarization filter. This proposal implements therefore a one-to-one correspondence between the degeneracy configuration of the polarizer states (list of numbers of identical polarizers) and the $\mathcal{D}_{\{n_i\}}$ entanglement family of the produced atomic state. The proof of this correspondence given in Ref.~\cite{Bas09} for $N=3$ is thus here straightforwardly extended to the general $N$ case. We think that the proposed general entanglement classification of symmetric $N$-qubit states, and its operational interpretation, will boost the related fields in quantum information, with theory and experiments walking together.

L. L. thanks the Alexander von Humboldt Foundation for funding. E. S. acknowledges UPV-EHU Grant
No. GIU07/40 and the EuroSQIP European project. T. B., S. K., and M. G. thank the Belgian FRS-FNRS.


\begin{thebibliography}{99}
\bibitem{footnote1} In number theory, a partition of a positive integer $N$ is a way of writing $N$ as a sum of positive integers. Two sums which only differ in the order of their summands are considered to be the same partition. The number of partitions of $N$ is given by the partition function $p(N)$.
\end{thebibliography}
\end{document}